\documentclass[a4paper]{article}
\pdfoutput=1

\usepackage[a4paper]{geometry}
\usepackage{microtype}

\usepackage{amsfonts}
\usepackage{amsmath}
\usepackage{amssymb}
\usepackage{latexsym}
\usepackage{bbold}
\usepackage{mathtools}
\usepackage{colonequals}

\usepackage{xspace}

\usepackage{float}
\usepackage{multirow}
\usepackage{multicol}
\usepackage{url}
\usepackage{xcolor}

\usepackage[noconfig]{ntheorem}
\newtheorem{theorem}{Theorem}[section]

\newtheorem{proposition}[theorem]{Proposition} 
\newtheorem{definition}[theorem]{Definition}
\newtheorem{claim}{Claim}[theorem]

\theorembodyfont{}
\newtheorem{example}[theorem]{Example}

\theoremstyle{nonumberplain}
\theoremheaderfont{\scshape}
\theorembodyfont{\normalfont}
\theoremsymbol{\ensuremath{_\blacksquare}}

\usepackage{tikz}
\usetikzlibrary{automata}
\usetikzlibrary{arrows}
\usetikzlibrary{trees}
\usetikzlibrary{fit}
\usetikzlibrary{positioning}

\renewcommand{\paragraph}[1]{\par\vspace{7pt}\noindent{\sffamily\bfseries #1.}}

\newcommand{\M}{\ensuremath{\mathsf{M}}\xspace}

\newcommand{\I}{\ensuremath{\mathsf{I}}\xspace}

\newcommand{\GOne}{\ensuremath{\mathsf{G_1}}\xspace}

\newcommand{\ShEx}{\ensuremath{\mathsf{ShEx}}\xspace}
\newcommand{\ShExZero}{\ensuremath{\mathsf{ShEx_0}}\xspace}

\newcommand{\dom}{\ensuremath{\mathsf{dom}}\xspace}
\newcommand{\ran}{\ensuremath{\mathsf{ran}}\xspace}

\newcommand{\interval}[1]{{\ensuremath{\mathord{\text{\normalfont\fontfamily{lmtt}\selectfont{}#1}}}}}
\newcommand{\NONE}{\interval{0}\xspace}
\newcommand{\ONE}{\interval{1}\xspace}
\newcommand{\MAYBE}{\interval{?}\xspace}
\newcommand{\MANY}{\interval{*}\xspace}
\newcommand{\PLUS}{\interval{+}\xspace}

\newcommand{\fit}{\ensuremath{\mathsf{fit}}\xspace}

\newcommand{\str}{\ensuremath{\mathsf{str}}}
\newcommand{\dbl}{{\,\mathord{:}\,}}

\newcommand{\arity}{\ensuremath{\mathsf{arity}}\xspace}
\newcommand{\source}{\ensuremath{\mathsf{source}}\xspace}
\newcommand{\target}{\ensuremath{\mathsf{target}}\xspace}
\newcommand{\lab}{\ensuremath{\mathsf{lab}}\xspace}
\newcommand{\out}{\ensuremath{\mathsf{out}}\xspace}

\newcommand{\typing}{\ensuremath{\mathit{typing}}\xspace}

\newcommand{\typedlearner}{\ensuremath{\mathsf{typed\text{-}learner}}\xspace}

\newcommand{\minus}{\mathbin{\setminus}}
\newcommand{\Typesets}{\ensuremath{\mathit{Typesets}}\xspace}
\newcommand{\Blue}[1]{{\color{blue!60!black}#1}}
\newcommand{\cover}{\ensuremath{\mathit{cover}}\xspace}

\def\clap#1{\hbox to 0pt{\hss#1\hss}}

\newcommand{\minarity}{\ensuremath{\mathsf{minarity}}\xspace}
\newcommand{\maxarity}{\ensuremath{\mathsf{maxarity}}\xspace}
\newcommand{\occur}{\ensuremath{\mathsf{occur}}\xspace}
\newcommand{\minoccur}{\ensuremath{\mathsf{minoccur}}\xspace}
\newcommand{\maxoccur}{\ensuremath{\mathsf{maxoccur}}\xspace}
\newcommand{\Can}{\ensuremath{\mathit{Can}}\xspace}

\def\qed {{                
   \parfillskip=0pt        
   \widowpenalty=10000     
   \displaywidowpenalty=10000  
   \finalhyphendemerits=0  
   \leavevmode             
   \unskip                 
   \nobreak                
   \hfil                   
   \penalty50              
   \hskip.2em              
   \null                   
   \hfill                  
   {$\Box$}
   \par}}                  

\floatstyle{ruled}
\newfloat{float@algorithm}{htb}{loa}
\floatname{float@algorithm}{Algorithm}
\newcounter{LineCounter@algorithm} 

\newenvironment{BasicCommands@algorithm}{%
}{%
}

\newenvironment{algorithm*}%
{%
\begin{BasicCommands@algorithm}%
\list{}{\itemindent 0em%
        \listparindent\itemindent
        \rightmargin  \leftmargin}%
\item\relax
}{%
\endlist
\end{BasicCommands@algorithm}%
}

{%
  \newcommand{\ResetLineCounter}{%
    \setcounter{LineCounter@algorithm}{0}%
  }%
  \begin{BasicCommands@algorithm}%
  \begin{float@algorithm}%
    \ResetLineCounter%
}%
{%
  \end{float@algorithm}%
  \end{BasicCommands@algorithm}%
}

\title{Inference of Shape Expression Schemas \\from Typed RDF Graphs}
\author{
Beno\^{i}t Groz\\
University Paris Sud\\
{Paris, France}
\and
Aurélien Lemay\\
University of Lille\\
Lille, France
\and
S\l{}awek Staworko\\
University of Lille\\
Lille, France
\and
Piotr Wieczorek\\
University of Wroc\l{}aw\\
Wroc\l{}aw, Poland
}

\begin{document}

\maketitle

\begin{abstract}
  We consider the problem of constructing a Shape Expression Schema (\ShEx) that
  describes the structure of a given input RDF graph. We employ the framework of
  \emph{grammatical inference}, where the objective is to find an inference
  algorithm that is both \emph{sound} i.e., always producing a schema that
  validates the input RDF graph, and \emph{complete} i.e., able to produce any
  schema, within a given class of schemas, provided that a sufficiently
  informative input graph is presented. We study the case where the input graph
  is \emph{typed} i.e., every node is given with its types. We limit our
  attention to a practical fragment \ShExZero of Shape Expressions Schemas that
  has an equivalent graphical representation in the form of \emph{shape
    graphs}. We investigate the problem of constructing a canonical
  representative of a given shape graph. Finally, we present a sound and
  complete algorithm for shape graphs thus showing that \ShExZero is learnable
  from typed graphs.
\end{abstract}

\section{Introduction}
\label{sec:intro}
Traditionally, in relational databases defining the schema is the mandatory
first step before a database can be even populated with data. Novel database
models, such as NoSQL and graph databases, quite intentionally allow to store
and process data without declaring any schema in order not to hinder the natural
evolution of the database structure while the applications around it are being
developed. In fact, often a suitable schema formalism is proposed long after a
particular database model has established its place in practice.  In those
circumstances a natural problem of \emph{schema inference} arises: given an
schema-less database construct a schema that captures the structure of the
database. This problem has been identified as an important research
direction~\cite{Dagstuhl16151} and is well motivated since the knowledge of
database structure is instrumental in any meaningful data processing tasks such
as querying or transformation.

\begin{figure}[htb]
  \centering
  \begin{tikzpicture}[>=latex,scale=1.25]
    \small
    \path[use as bounding box] (-7.5,-5) rectangle (2.5,1.8);

    \begin{scope}[xshift=-0.65cm]
    \node (bug1) at (-1,1) {\tt bug${}_1$};
    \node (bug2) at (2.5,1) {\tt bug${}_2$};
    \node (bug3) at (-3,0) {\tt bug${}_3$};
    \node (bug4) at (1,0) {\tt bug${}_4$};
    \node (user1) at (-1.75,-1.75) {\tt user${}_1$};
    \node (user2) at (2.75,-1) {\tt user${}_2$};
    \node (emp1) at (-.2,-1.75) {\tt emp${}_1$};

    \node (d1) at (-2.95,1.45) {\it ``Boom!''};
    \node (d2) at (0.5,1.45) {\it ``Kaboom!''};
    \node (d3) at (-3.1,-1.4) {\it ``Kabang!''};
    \node (d4) at (1.25,-1.5) {\it ``Bang!''};

    \node (n1) at (-3,-2.5) {\it ``John''};
    \node (n2) at (-1.1,-2.75) {\it ``Mary''};
    \node (e2) at (0.7,-2.75) {\it ``m@h.org''};
    \node (n3) at (2,-2) {\it ``Steve''};

    \draw (user1) edge[->] node[above,sloped] {\tt name} (n1);

    \draw (user2) edge[->] node[above,sloped] {\tt name} (n3);

    \draw (emp1) edge[->] node[above,sloped] {\tt name} (n2);
    \draw (emp1) edge[->] node[above,sloped] {\tt email} (e2);    

    \draw (bug1) edge[->] node[above,sloped] {\tt related} (bug4);
    \draw (bug1) edge[->] node[above,sloped] {\tt related} (bug3);
    \draw (bug1) edge[->] node[above,sloped] {\tt submittedBy} (user1);
    \draw (bug1) edge[->] node[above,sloped] {\tt verifiedBy} (emp1);
    \draw (bug1) edge[->] node[above,sloped] {\tt descr} (d1);

    \draw (bug2) edge[->] node[above,sloped] {\tt related} (bug4);
    \draw (bug2) edge[->] node[above,sloped] {\tt submittedBy} (user2);
    \draw (bug2) edge[->] node[above,sloped] {\tt descr} (d2);

    \draw (bug3) edge[->] node[above,sloped] {\tt submittedBy} (user1);
    \draw (bug3) edge[->] node[above,sloped] {\tt descr} (d3);

    \draw (bug4) edge[->] node[below,sloped] {\tt submittedBy} (emp1);
    \draw (bug4) edge[->] node[above,sloped] {\tt descr} (d4);
    \end{scope}
  
    \begin{scope}[yshift=0.15cm]
      \node (Bug) at (-6.5, 0.5) {\sf Bug};
      \node (User) at (-7.5, -1.75) {\sf User};
      \node (Emp) at (-5.25, -1.75) {\sf Employee};
      \node (s) at (-6.5, -3) {$\str$};

      \draw[loop]  (Bug) edge[->] 
                         node[above] {\tt related} 
                         node[below] {\MANY} (Bug); 

      \draw (Bug) edge[->] 
                  node[above,sloped] {\tt submittedBy} 
                  node[below,sloped] {\ONE}
                  (User);
      \draw (Bug) edge[->] 
                  node[above,sloped] {\tt verifiedBy} 
                  node[below,sloped] {\MAYBE}
                  (Emp);

      \draw[bend angle=4] (Bug) edge[->,bend left]
                  node[above,sloped] {\tt descr}
                  node[below,sloped] {\ONE}
                  (s);

      \draw[bend angle=25] (User) edge[->, bend left] 
                   node[above,sloped,pos=0.35] {\tt name} 
                   node[below,sloped] {\ONE}
                   (s);

      \draw[bend angle=45] (User) edge[->, bend right] 
                   node[above,sloped] {\tt email} 
                   node[below,sloped] {\MAYBE}
                   (s);

      \draw[bend angle=20] (Emp) edge[->, bend right] 
                   node[above,sloped,pos=0.4] {\tt name} 
                   node[below,sloped] {\ONE}
                   (s);

      \draw[bend angle=45] (Emp) edge[->, bend left] 
                   node[above,sloped] {\tt email} 
                   node[below,sloped] {\ONE}
                   (s);
                 
    \end{scope}
\node[anchor=center, text width=12.75cm] at (-2.625,-4) {
  \begin{align*}
    & \mathsf{Bug} \rightarrow 
    \mathtt{descr}\dbl\str ,\ 
    \mathtt{submittedBy}\dbl{}\mathsf{User} ,\ 
    \mathtt{verifiedBy}\dbl{}\mathsf{Employee}\MAYBE ,\ 
    \mathtt{related}\dbl{}\mathsf{Bug}\MANY\\
    & \mathsf{User} \rightarrow 
    \mathtt{name}\dbl\str ,\  
    \mathtt{email}\dbl{}\str\MAYBE\\
    & \mathsf{Employee} \rightarrow 
    \mathtt{name}\dbl\str ,\  
    \mathtt{email}\dbl{}\str
  \end{align*}
};

  \end{tikzpicture}
  \caption{An RDF graph with bug reports (top right) together with a
    shape expression schema (bottom) and the corresponding shape graph
    (top left). $\str$ is a built-in type for literal string nodes.}
  \label{fig:rdf-bug-reports}
\end{figure}
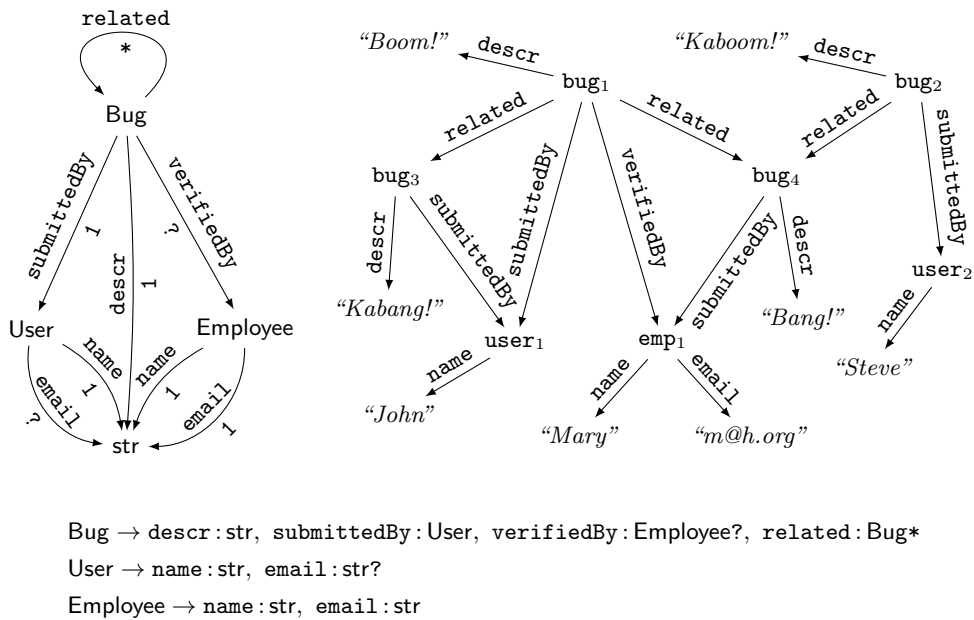

In the present paper, we present a principled approach to the problem of
inference of schema for graph databases. We consider RDF graphs and Shape
Expression Schemas (ShEx)~\cite{ShExW3C,SBLGHPS15}. ShEx builds on the success
of XML Schema and allows to describe the \emph{structure} of an RDF graph by
defining patterns of arrangement of RDF nodes. More precisely, ShEx specifies a
collection of node types, each type defined by a regular expression that
constrains the types of the outbound neighborhood of a node. Take for instance
the RDF graph storing bug reports, presented in Figure~\ref{fig:rdf-bug-reports}
together with its shape expression schema.
The schema requires a bug report to have a description and a user who submitted
it. Optionally, a bug report may have an employee who verified it. Also, a bug
report can have a number of related bug reports. A user has a name and an
optional email address while an employee has a name and a mandatory email
address. We point out that just like with XML Schema the nodes of the RDF graph
need not by typed and it is the task of a validation algorithm to find for a
valid node typing~\cite{PrGaSo15,GPSR15,SBLGHPS15}, and furthermore, some nodes
may need to have more than one type e.g., $\mathtt{emp}_1$ needs to have the
types $\mathsf{User}$ and $\mathsf{Employee}$.

We focus our investigation on a practical subclass \ShExZero that allows type
definitions with collections of atoms with multiplicities ranging over \ONE,
\MAYBE, \PLUS, and \MANY, and does not allow disjunction or grouping. \ShExZero
is particularly suited to capture the topology of RDF graphs obtained by
exporting relational databases in a number of formalisms proposed for this task,
such as R2RML, Direct Mapping, and YARRRML~\cite{SeArMi12,STCM11,BoLoSt18}.
Also, \ShExZero has a significant overlap with an alternative schema language
for RDF, the Shape Constraint Language (SHACL)~\cite{Boneva16}.  More
importantly, the class \ShExZero enjoys a useful and sought-after feature of
having an equivalent graphical representation in the form of a \emph{shape
  graph}, where nodes are types and edges are labeled by both a symbol and a
multiplicity (cf. Figure~\ref{fig:rdf-bug-reports}).

\begin{figure}[htb]
  \centering
  \begin{tikzpicture}[>=latex,scale=1.5]
    \small

    \begin{scope}[xshift=-0.65cm]
    \node (bug1) at (-1,1) {\tt bug${}_1$};
    \node[right=-.2cm of bug1] {\scriptsize\Blue{\sf:Bug}};
    \node (bug2) at (2.5,1) {\tt bug${}_2$};
    \node[right=-.2cm of bug2] {\scriptsize\Blue{\sf:Bug}};
    \node (bug3) at (-3,0) {\tt bug${}_3$};
    \node[right=-.2cm of bug3] {\scriptsize\Blue{\sf:Bug}};
    \node (bug4) at (1,0) {\tt bug${}_4$};
    \node[right=-.2cm of bug4] {\scriptsize\Blue{\sf:Bug}};
    \node (user1) at (-1.75,-1.75) {\tt user${}_1$};
    \node[right=-.2cm of user1] {\scriptsize\Blue{\sf:User}};
    \node (user2) at (2.75,-1) {\tt user${}_2$};
    \node[right=-.2cm of user2] {\scriptsize\Blue{\sf:User}};
    \node (emp1) at (-.2,-1.75) {\tt emp${}_1$};
    \node[right=-.2cm of emp1] {\scriptsize\Blue{\sf:Employee,\,User}};

    \node (d1) at (-2.95,1.45) {\it ``Boom!''};
    \node[right=-.2cm of d1] {\scriptsize\Blue{\sf:\str}};
    \node (d2) at (0.5,1.45) {\it ``Kaboom!''};
    \node[right=-.2cm of d2] {\scriptsize\Blue{\sf:\str}};
    \node (d3) at (-3.1,-1.4) {\it ``Kabang!''};
    \node[right=-.2cm of d3] {\scriptsize\Blue{\sf:\str}};
    \node (d4) at (1.25,-1.5) {\it ``Bang!''};
    \node[right=-.2cm of d4] {\scriptsize\Blue{\sf:\str}};

    \node (n1) at (-3,-2.5) {\it ``John''};
    \node[right=-.2cm of n1] {\scriptsize\Blue{\sf:\str}};
    \node (n2) at (-1.1,-2.75) {\it ``Mary''};
    \node[right=-.2cm of n2] {\scriptsize\Blue{\sf:\str}};
    \node (e2) at (0.7,-2.75) {\it ``m@h.org''};
    \node[right=-.2cm of e2] {\scriptsize\Blue{\sf:\str}};
    \node (n3) at (2,-2) {\it ``Steve''};
    \node[right=-.2cm of n3] {\scriptsize\Blue{\sf:\str}};

    \draw (user1) edge[->] node[above,sloped] {\tt name} (n1);

    \draw (user2) edge[->] node[above,sloped] {\tt name} (n3);

    \draw (emp1) edge[->] node[above,sloped] {\tt name} (n2);
    \draw (emp1) edge[->] node[above,sloped] {\tt email} (e2);    

    \draw (bug1) edge[->] node[above,sloped] {\tt related} (bug4);
    \draw (bug1) edge[->] node[above,sloped] {\tt related} (bug3);
    \draw (bug1) edge[->] node[above,sloped] {\tt submittedBy} (user1);
    \draw (bug1) edge[->] node[above,sloped] {\tt verifiedBy} (emp1);
    \draw (bug1) edge[->] node[above,sloped] {\tt descr} (d1);

    \draw (bug2) edge[->] node[above,sloped] {\tt related} (bug4);
    \draw (bug2) edge[->] node[above,sloped] {\tt submittedBy} (user2);
    \draw (bug2) edge[->] node[above,sloped] {\tt descr} (d2);

    \draw (bug3) edge[->] node[above,sloped] {\tt submittedBy} (user1);
    \draw (bug3) edge[->] node[above,sloped] {\tt descr} (d3);

    \draw (bug4) edge[->] node[below,sloped] {\tt submittedBy} (emp1);
    \draw (bug4) edge[->] node[above,sloped] {\tt descr} (d4);
    \end{scope}
  \end{tikzpicture}
  \caption{An typed RDF graph with bug reports. Types are in blue.}
  \label{fig:rdf-bug-reports-typed}
\end{figure}
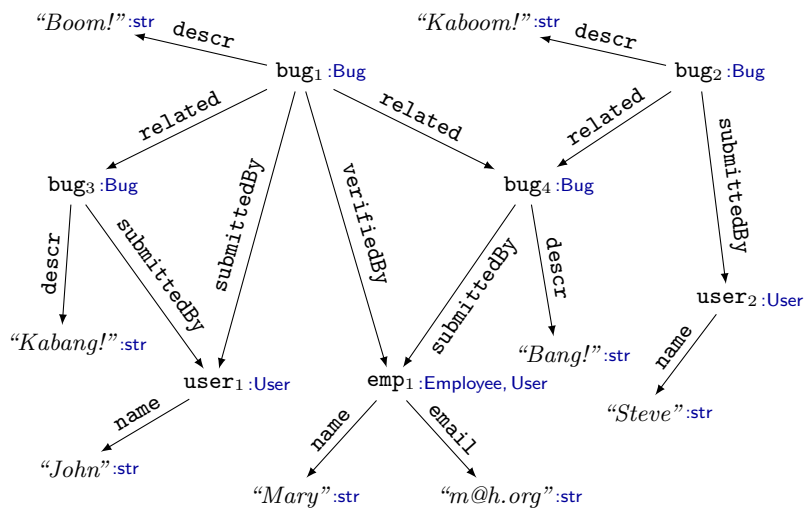

In this paper, we present our findings on learning shape graphs from \emph{typed
  graphs}, graphs whose nodes come with type informations. Indeed, in RDF there
is the predicate \texttt{rdf:type}, which is the dedicated element of standard
vocabulary intended for the purpose of assigning types to nodes of the graph. In
out work we assume that the \emph{typing} that assigns to every node a set of
its types, is given in a manner independent of the graph. As an example the
typed version of the graph in Figure~\ref{fig:rdf-bug-reports} is presented in
Figure~\ref{fig:rdf-bug-reports-typed}.
Consequently, the problem of schema inference is reduced to inferring type
definitions. This problem may seem trivial because in fact it is simple to solve
when nodes have precisely one type. For instance, if we consider in
Figure~\ref{fig:rdf-bug-reports-typed} all nodes that have type $\mathsf{User}$,
then the definition of this type is straightforward since all nodes have an
outgoing $\mathtt{name}$ edge leading to a node whose type is precisely $\str$
and some nodes have an outgoing $\mathtt{email}$ edge that also leads to a node
whose type is unambiguously $\str$. Hence, the type definition
$\mathsf{User} \rightarrow \mathtt{name}\dbl\str ,\
\mathtt{email}\dbl{}\str\MAYBE$.  

The inference of type definition is, however, less obvious when the nodes have
multiple types: when inferring type definition we need to make choice of the
relevant type. For instance consider the nodes of type $\mathsf{Bug}$ and notice
that $\mathtt{bug}_1$ has an outgoing edge $\mathtt{verifiedBy}$ that leads to a
node $\mathtt{emp}_1$ that has both types $\mathsf{User}$ and
$\mathsf{Employee}$. Depending on the choice the definition of $\mathsf{Bug}$
can become either
\[
  \mathsf{Bug} \rightarrow 
    \mathtt{descr}\dbl\str ,\ 
    \mathtt{submittedBy}\dbl{}\mathsf{User} ,\ 
    \mathtt{verifiedBy}\dbl{}\mathsf{User}\MAYBE ,\ 
    \mathtt{related}\dbl{}\mathsf{Bug}\MANY
\]
or 
\[
  \mathsf{Bug} \rightarrow 
    \mathtt{descr}\dbl\str ,\ 
    \mathtt{submittedBy}\dbl{}\mathsf{User} ,\ 
    \mathtt{verifiedBy}\dbl{}\mathsf{Employee}\MAYBE ,\ 
    \mathtt{related}\dbl{}\mathsf{Bug}\MANY\ .
\]
Naturally, these two definitions are not equivalent and the question is which
one should be chosen. We find that the second type definition is more
appropriate for two reasons. First, we observe that
$\mathsf{Employee}\subseteq\mathsf{User}$ i.e., every node that has the type
$\mathsf{Employee}$ will also have the type $\mathsf{User}$. Indeed, this can be
established by observing the typed graph alone: there is no node that has type
$\mathsf{Employee}$ and not $\mathsf{User}$ (while the converse is true: there
are node that have type $\mathsf{User}$ but not $\mathsf{Employee}$). Secondly,
should the type $\mathsf{User}$ be used in the context of the edge
$\mathtt{verifiedBy}$, then we should be able to find an edge
$\mathtt{verifiedBy}$ that leads to a node that has only the type
$\mathsf{User}$. In general, our algorithm for constructing the appropriate type
definitions is based on a comprehensive analysis of the typing information
present in the input typed graph.

More importantly, our approach is a solution to the inference problem stated
with the use of \emph{grammatical inference} framework~\cite{Gold78}, which in
recent years has been successfully applied to a number of database formalisms
ranging from queries~\cite{CGLN07,StWi12} to schemas~\cite{BeGeNeVa10,CiSt13} to
transformations~\cite{LeMaNi10,LLNST14}. In essence, an inference algorithm
needs to be both \emph{sound} i.e., producing a schema that validates the input
graph, and \emph{complete} i.e., able to infer any goal schema with a
sufficiently informative input graph, typically referred to as
\emph{characteristic} graph of the goal language of typed graphs. 

In this paper, we present an algorithm that is both sound and complete for the
full class of shape graphs (\ShExZero). Interestingly, our investigations into
learnability lead us to study the problem of \emph{canonization}. Namely, for a
given shape graph there might be a large number of equivalent shape graphs that
define the same language of typed graphs. When presented with a characteristic
graph, the inference algorithm outputs one of the shape graphs that define the
goal language. It is desirable for the algorithm to output the same shape graph
regardless of how the characteristic graph is constructed and returning the
canonical representative is an elegant approach to address this
need. Consequently, we present an effective characterization of canonical shape
graphs based on a canonization procedure and design our inference algorithm in
such a way that it too returns the canonical shape graph that defines the goal
language.

This paper is organized ass follows. In Section~\ref{sec:basics} we present
basic notions. In Section~\ref{sec:framework} we define the framework of
grammatical inference for shape graphs. In Section~\ref{sec:typed-graphs} we
present basic instruments for analyzing the input typed graph. In
Section~\ref{sec:canonization} we investigate the problem of canonization of
shape graphs. In Section~\ref{sec:infererence} we present a sound and complete
inference algorithm for shape graphs from typed graphs. In
Section~\ref{sec:related} we discuss the related work. Finally, in
Section~\ref{sec:conclusions} we summarize our findings and outline future
directions of study.

\section{Basic notions}
\label{sec:basics}

Throughout this paper we employ elements of function notation to relations. For
instance, for a binary relation $R\subseteq A\times B$ we set
$\dom(R)=\{a\in A \mid \exists b\in B.\ (a,b)\in R\}$,
$\ran(R)=\{b\in B \mid \exists a\in A.\ (a,b)\in R\}$,
$R(a) = \{b\in B \mid (a,b)\in R\}$ for $a\in A$, and
$R^{-1}(b)=\{a\in A \mid (a,b)\in R\}$ for $b\in B$.

\paragraph{Intervals} 
We use the standard notation $[n;m]$ to denote \emph{intervals}, which represent
nonempty sets of consecutive natural numbers with $0\leq n\leq m\leq \infty$.
The value $n$ (resp. $m$) is called the minimum of the interval (resp. the
maximum). By $\I$ we denote the set of all intervals although our schema
formalisms use only \emph{basic intervals}
$\M=\{\NONE,\MAYBE,\ONE,\MANY,\PLUS\}$ for which we employ a shorthand notation:
$\NONE$ is $[0;0]$, $\MAYBE$ is $[0;1]$, $\ONE$ is $[1;1]$, $\PLUS$ is
$[1;\infty]$, and $\MANY$ is $[0;\infty]$.

We employ the point-wise addition operation
$[n_1;m_1]\oplus[n_2;m_2]=[n_1+n_2;m_1+m_2]$ and the natural interpretation of
the inclusion relation $[n_1;m_1]\subseteq[n_2;m_2]$ iff $n_2\leq n_1$ and
$m_1\leq m_2$. Finally, we define the function $\fit$ that maps any set $X$ of
natural numbers (occurrences) into a smallest interval in $\I$ that contains all
elements of $X$.

\paragraph{Graphs}
We assume a fixed and finite set $\Sigma$ of edge labels and a fixed and finite
set of types $\Gamma$.

\begin{definition}[Graph]
\label{def:graph}
A \emph{graph} $G$ is a pair $(N_G,E_G)$, where $N_G$ is a set of nodes and
$E_G\subseteq N_G\times\Sigma\times N_G$ is a set of oriented labeled
edges. \qed
\end{definition}
For an edge $e=(n,a,m)$ we set $\source(e)=n$, $\lab(e)=a$, and
$\target(e)=m$. Also, for a node $n$ of a graph $G$ we identify its set of
\emph{outbound edges} $\out_G(n)=\{e\in E_G \mid \source(e)=n\}$.

\begin{figure}[htb]
  \begin{minipage}{0.30\textwidth}%
    \centering
    \begin{tikzpicture}[>=latex]
    \footnotesize
    
    \begin{scope}[xshift=0cm,bend angle=15]
    \node at (2,1.5) {~};
    \node (bug3) at (0,0) {\tt b${}_3$};
    \node (bug1) at (1,1) {\tt b${}_1$};
    \node (bug4) at (2.25,0) {\tt b${}_4$};
    \node (bug2) at (3.0125,1) {\tt b${}_2$};

    \node (user1) at (0,-1.25) {\tt u${}_1$};
    \node (user2) at (3.125,-1.25) {\tt u${}_2$};
    \node (emp1) at (1.25,-1.025) {\tt e${}_1$};

    \node (s) at (1,-2.5) {$\bullet$};

    \draw (bug1) edge[->,bend left] node[above,sloped] {\tt r} (bug4);
    \draw (bug1) edge[->,bend right] node[above,sloped] {\tt r} (bug3);
    \draw (bug1) edge[->,bend left] node[above,sloped] {\tt s} (user1);
    \draw (bug1) edge[->,bend right] node[above,sloped] {\tt v} (emp1);

    \draw[bend angle=05] (bug2) edge[->,bend right] node[above,sloped] {\tt r} (bug4);
    \draw[bend angle=05] (bug2) edge[->,bend left] node[above=-1pt,sloped] {\tt s} (user2);

    \draw (bug3) edge[->,bend right] node[above,sloped] {\tt s} (user1);

    \draw[bend angle=05] (bug4) edge[->,bend right] node[above=-1pt,sloped] {\tt s} (emp1);

    \draw[bend angle=09.25] (bug1) edge[->, bend right] node[below=-1pt,sloped,pos=0.6] {\tt d} (s);
    \draw (bug2) edge[->, bend left,in=137,out=20] node[above=-1pt,sloped,pos=0.35] {\tt d} (s);
    \draw (bug3) edge[->, bend right,out=-80,in=-110] node[below=-1pt,sloped] {\tt d} (s);
    \draw (bug4) edge[->, bend left,in=140] node[above=-1pt,sloped,pos=0.45] {\tt d} (s);
    \draw (user1) edge[->, bend right] node[above,sloped] {\tt n} (s);
    \draw[bend angle=20] (user2) edge[->,bend left] node[above=-1pt,sloped,pos=0.3] {\tt n} (s);
    \draw[bend angle=10] (emp1) edge[->,bend right] node[below=-1pt,sloped,pos=0.4] {\tt n} (s);
    \draw[bend angle=25] (emp1) edge[->,bend left] node[below=-1pt,sloped] {\tt e} (s);    
    \end{scope}
    \end{tikzpicture}
    \caption{Simple graph.}
    \label{fig:graphs}
  \end{minipage}
  \hfill
  \begin{minipage}{0.30\textwidth}
    \centering
    \begin{tikzpicture}[>=latex]
    \footnotesize
    \begin{scope}[xshift=-2.5cm]
      \node at (-0.5,1.5) {$S_0$};
      \node (Bug) at (0, 0.5) {\color{blue!75!black}\sf B};
      \node (User) at (-.75, -1) {\color{blue!75!black}\sf U};
      \node (Emp) at (.75, -1) {\color{blue!75!black}\sf E};
      \node (str) at (0, -2.5) {\color{blue!75!black}$\str$};

      \draw[loop]  (Bug) edge[->,in=120,out=60,looseness=5] 
                         node[above] {$\mathtt{r}\MANY$} (Bug); 

      \draw (Bug) edge[->] 
                  node[above,sloped] {$\mathtt{s}$}
                  (User);
      \draw (Bug) edge[->] 
                  node[above,sloped] {$\mathtt{v}\MAYBE$}
                  (Emp);

      \draw[bend angle=0.1] (Bug) edge[->,bend left]
                  node[above=-1.5pt,sloped] {$\mathtt{d}$}
                  (str);

      \draw[bend angle=15] (User) edge[->, bend right] 
                   node[above,sloped,pos=0.5] {$\mathtt{n}$}
                   (str);

      \draw[bend angle=55] (User) edge[->, bend right] 
                   node[above=-1.5pt,sloped] {$\mathtt{e}\MAYBE$}
                   (str);

      \draw[bend angle=15] (Emp) edge[->, bend left] 
                   node[above,sloped,pos=0.5] {$\mathtt{n}$}
                   (str);

      \draw[bend angle=55] (Emp) edge[->, bend left] 
                   node[above=-1.5pt,sloped] {$\mathtt{e}$}
                   (str);
                 
    \end{scope}
    \end{tikzpicture}
    \caption{\mbox{Shape graph $S_0$.}}
    \label{fig:graphs}
    \end{minipage}
    \hfill
    \begin{minipage}{0.30\textwidth}
    \centering
    \begin{tikzpicture}[>=latex]
      \footnotesize

    \begin{scope}[xshift=0cm,bend angle=15]
    \node at (2,1.5) {$G_0$};
    \node (bug3) at (0,0) {\tt b${}_3$};
    \node[right=-.25cm of bug3] {\Blue{\sf:B}};
    \node (bug1) at (1,1) {\tt b${}_1$};
    \node[right=-.25cm of bug1] {\Blue{\sf:B}};
    \node (bug4) at (2.25,0) {\tt b${}_4$};
    \node[right=-.25cm of bug4] {\Blue{\sf:B}};
    \node (bug2) at (3.0125,1) {\tt b${}_2$};
    \node[right=-.25cm of bug2] {\Blue{\sf:B}};

    \node (user1) at (0,-1.25) {\tt u${}_1$};
    \node[right=-.25cm of user1] {\Blue{\sf:U}};
    \node (user2) at (3.125,-1.25) {\tt u${}_2$};
    \node[right=-.25cm of user2] {\Blue{\sf:U}};
    \node (emp1) at (1.25,-1.025) {\tt e${}_1$};
    \node[right=-.25cm of emp1] {\Blue{\sf:U,E}};

    \node (s) at (1,-2.5) {$\bullet$};
    \node[below=-.125cm of s] {\Blue{\sf:str}};

    \draw (bug1) edge[->,bend left] node[above,sloped] {\tt r} (bug4);
    \draw (bug1) edge[->,bend right] node[above,sloped] {\tt r} (bug3);
    \draw (bug1) edge[->,bend left] node[above,sloped] {\tt s} (user1);
    \draw (bug1) edge[->,bend right] node[above,sloped] {\tt v} (emp1);

    \draw[bend angle=05] (bug2) edge[->,bend right] node[above,sloped] {\tt r} (bug4);
    \draw[bend angle=05] (bug2) edge[->,bend left] node[above=-1pt,sloped] {\tt s} (user2);

    \draw (bug3) edge[->,bend right] node[above,sloped] {\tt s} (user1);

    \draw[bend angle=05] (bug4) edge[->,bend right] node[above=-1pt,sloped] {\tt s} (emp1);

    \draw[bend angle=09.25] (bug1) edge[->, bend right] node[below=-1pt,sloped,pos=0.6] {\tt d} (s);
    \draw (bug2) edge[->, bend left,in=137,out=20] node[above=-1pt,sloped,pos=0.35] {\tt d} (s);
    \draw (bug3) edge[->, bend right,out=-80,in=-110] node[below=-1pt,sloped] {\tt d} (s);
    \draw (bug4) edge[->, bend left,in=140] node[above=-1pt,sloped,pos=0.45] {\tt d} (s);
    \draw (user1) edge[->, bend right] node[above,sloped] {\tt n} (s);
    \draw[bend angle=20] (user2) edge[->,bend left] node[above=-1pt,sloped,pos=0.3] {\tt n} (s);
    \draw[bend angle=10] (emp1) edge[->,bend right] node[below=-1pt,sloped,pos=0.4] {\tt n} (s);
    \draw[bend angle=25] (emp1) edge[->,bend left] node[below=-1pt,sloped] {\tt e} (s);    
    \end{scope}
    \end{tikzpicture}
    \caption{Typed graph $G_0$.}
    \label{fig:graphs}
  \end{minipage}
\end{figure}

\begin{definition}[Shape graph]
  A \emph{shape graph} $S$ is a function
  $\arity_S: \Gamma\times\Sigma\times\Gamma\rightarrow\M$ decorates with basic
  intervals the edges of the complete graph whose nodes are types $\Gamma$.  By
  $\ShExZero$ we denote the set of all shape graphs.
\end{definition}
The semantics of shape graphs is defined with the notion of typings that
associate to nodes of a graph set of types. 
\begin{definition}[Typing]
\label{def:typing}
A node $n\in N_G$ of a graph $G$ \emph{satisfies} type $t$ w.r.t.\ a shape graph
$S$ iff there is a witness $\lambda :\out_G(n)\rightarrow \Gamma$ such that the
following conditions are satisfied:
\begin{enumerate}
\item for every outbound edge $e\in\out_G(n)$ leads to a node $\target(e)$ that
  satisfies $\lambda(e)$, and 
\item for every $a\in\Sigma$ and every $s\in\Gamma$ the number of $a$-labeled
  edges outgoing from $n$ that are assigned by $\lambda$ the type $s$ is
  contained by the interval $\arity_S(t,a,s)$, or formally,
  $|\{e\in\out_G(n)\mid\lab(e)=a,\ \lambda(e)=s\}|\in \arity_S(t,a,s)$.
\end{enumerate}
The \emph{typing} of a graph $G$ w.r.t.\ $S$ is the relation
$\typing\subseteq N_G\times \Gamma$ such that $(n,t)\in\typing$ iff $n$
satisfies $t$ w.r.t.\ $S$. The typing is \emph{proper} iff
$\typing(n)\neq\emptyset$ for every $n\in N_G$.
\end{definition}
In this paper, we work only with graphs that are typed w.r.t.\ some shape graph.
\begin{definition}[Typed graph]
  Given a shape graph $S$, a \emph{typed graph} $G=(N_G,E_G,\typing_G)$ is a
  graph extended with a proper typing (w.r.t. a shape graph). The language
  $L(S)$ defined by a shape graph $S$ is the set of all finite graphs typed with
  a proper typing w.r.t. $S$. By $\GOne$ we denote the set of all finite typed
  graphs $\GOne=\bigcup \{L(S)\mid S\in\ShExZero\}$.
\end{definition}
Two shape graphs $S_1$ and $S_2$ are \emph{equivalent}, in symbols
$S_1\equiv S_2$, iff $L(S_1)=L(S_2)$. Given two typed graphs $G$ and $G'$ their
disjoint union $G\uplus G'$ is
$(N_G\uplus N_{G'},E_G\uplus E_{G'},\typing_G\uplus\typing_{G'})$.

\paragraph{Type expressions and equations}
Throughout this paper, we employ expressions that use set operators on types in
$\Gamma$. These expressions have a straightforward meaning a straightforward
interpretation in the context of a fixed shape graph $S$: a type name $t$ is
simply replaced by the set $\typing_{\uplus L(S)}(t)$ of all nodes that satisfy
it. For instance, $t_1=t_2$ means that the types $t_1$ and $t_2$ are
\emph{equivalent} i.e., a node satisfies either both or neither. The expression
$t \subseteq t_1 \cup t_2$ reads as ``any node with type $t$ has also type $t_1$
or type $t_2$.''


\section{Learning framework}
\label{sec:framework}
We employ the framework of grammatical inference \cite{Gold78} to inference of
\ShEx schemas from typed graphs. In essence, this framework require the
existence of a learning algorithm capable of inferring every schema from a
sufficiently informative input typed graph. Such a typed graph is called
\emph{characteristic} and to avoid collusion the inference algorithm is required
be conservative: it must infer the goal schema even if to the characteristic
graph we add other potentially less informative fragments. Formally, $G$
\emph{extends} $G'$ \emph{consistently with schema} $S$ iff there is $G''$ such
that $G=G'\uplus G''$ and $G,G',G''\in L(S)$.
\begin{definition}
  \label{def:inference}
  Shape graphs are \emph{learnable from typed graphs in polynomial time} iff
  there is a polynomial inference algorithm $\mathsf{learner}$ such that
  \begin{description}
  \item[Soundness] For every input typed graph $G\in\GOne$ the inference
    algorithm returns a shape graph $\mathsf{learner}(G)=S$ such that
    $G\in L(S)$.
  \item[Completeness] For every shape graph $S\in\ShExZero$ there exists a
    \emph{characteristic graph} $G \in L(S)$ such that for any $G'\in L(S)$ we
    have $\mathsf{learner}(G\uplus G')\equiv S$.  \qed
  \end{description}
\end{definition}


\section{Typed graphs}
\label{sec:typed-graphs}
In this section we introduce tools for inspecting typed graphs and extracting
the relevant typing information for inference algorithm. Throughout this section
we fix a shape graphs $S$.

\paragraph{Contexts and type definition fragments}
As we illustrate next, when inferring a type definition, we only need to inspect
the local outbound neighborhood of nodes of the type in question, we can ignore
the identity of nodes and focus on types alone. More importantly, the definition
of type can be inferred in fragments independently for each outgoing edge label.
\begin{example}
  \label{ex:contexts}
  Take the graph $G_0$ in Figure~\ref{fig:example} and the typing corresponding
  to the presented embedding of $G_0$ in $S_0$. 
  \begin{figure}[htb]
  \label{fig:example}
    \centering
    \begin{tikzpicture}[>=latex,
      punkt/.style={circle,minimum size=0.1cm,draw,fill,inner sep=0pt,outer sep=0.125cm}
      ]
      \footnotesize
      \begin{scope}[xshift=-.25cm,yshift=-.125cm,gray]
        \node[left] at (0,-0.5) {$\mathtt{u}_1$};
        \node[left] at (0,-1) {$\mathtt{u}_2$};
        \node[left] at (0,-1.5) {$\mathtt{e}_1$};
      \end{scope}
      \begin{scope}[xshift=0cm]
        \node[right] at (0,0.125) {$C_0=(\mathsf{U},\mathtt{n})$};
        \begin{scope}[yshift=-0.125cm]
        \node[punkt] (x) at (0.15,-0.5) {};
        \node[punkt] (y) at (1,-0.5) {} edge[<-] (x);
        \node[right] at (1.05,-0.5) {$\{\str\}$};
        \node[punkt] (x) at (0.15,-1) {};
        \node[punkt] (y) at (1,-1) {} edge[<-] (x);
        \node[right] at (1.05,-1) {$\{\str\}$};
        \node[punkt] (x) at (0.15,-1.5) {};
        \node[punkt] (y) at (1,-1.5) {} edge[<-] (x);
        \node[right] at (1.05,-1.5) {$\{\str\}$};
        \end{scope}        
      \end{scope}
      \begin{scope}[xshift=2.25cm]
        \node[right] at (0,0.125) {$C_1=(\mathsf{U},\mathtt{e})$};
        \begin{scope}[yshift=-0.125cm]
        \node[punkt] (x) at (0.15,-0.5) {};
        \node[punkt] (x) at (0.15,-1) {};
        \node[punkt] (y) at (1,-1) {} edge[<-] (x);
        \node[right] at (1.05,-1) {$\{\str\}$};
        \node[punkt] (x) at (0.15,-1.5) {};
        \node[punkt] (y) at (1,-1.5) {} edge[<-] (x);
        \node[right] at (1.05,-1.5) {$\{\str\}$};
        \end{scope}
      \end{scope}
      \begin{scope}[xshift=5.75cm]
      \begin{scope}[xshift=-.25cm,gray]
        \node[left] at (0,-0.5) {$\mathtt{b}_1$};
        \node[left] at (0,-1)   {$\mathtt{b}_2$};
        \node[left] at (0,-1.5) {$\mathtt{b}_3$};
        \node[left] at (0,-2) {$\mathtt{b}_4$};
      \end{scope}
      \begin{scope}[xshift=0cm]
        \node[right] at (0,0.125) {$C_2=(\mathsf{B},\mathtt{r})$};
        \node[punkt] (x) at (0.15,-0.5) {};
        \node[punkt] (y) at (1,-0.325) {} edge[<-] (x);
        \node[right] at (1.05,-0.325) {$\{\mathsf{B}\}$};
        \node[punkt] (y) at (1,-0.675) {} edge[<-] (x);
        \node[right] at (1.05,-0.675) {$\{\mathsf{B}\}$};
        \node[punkt] (x) at (0.15,-1) {};
        \node[punkt] (y) at (1,-1) {} edge[<-] (x);
        \node[right] at (1.05,-1) {$\{\mathsf{B}\}$};
        \node[punkt] (x) at (0.15,-1.5) {};
        \node[punkt] (x) at (0.15,-2) {};
      \end{scope}
      \begin{scope}[xshift=2.25cm]
        \node[right] at (0,0.125) {$C_3=(\mathsf{B},\mathtt{s})$};
        \node[punkt] (x) at (0.15,-0.5) {};
        \node[punkt] (y) at (1,-0.5) {} edge[<-] (x);
        \node[right] at (1.05,-0.5) {$\{\mathsf{U}\}$};
        \node[punkt] (x) at (0.15,-1) {};
        \node[punkt] (y) at (1,-1) {} edge[<-] (x);
        \node[right] at (1.05,-1) {$\{\mathsf{U}\}$};
        \node[punkt] (x) at (0.15,-1.5) {};
        \node[punkt] (y) at (1,-1.5) {} edge[<-] (x);
        \node[right] at (1.05,-1.5) {$\{\mathsf{U}\}$};
        \node[punkt] (x) at (0.15,-2) {};
        \node[punkt] (y) at (1,-2) {} edge[<-] (x);
        \node[right] at (1.05,-2) {$\{\mathsf{U},\mathsf{E}\}$};
      \end{scope}
      \begin{scope}[xshift=4.5cm]
        \node[right] at (0,0.125) {$C_4=(\mathsf{B},\mathtt{v})$};
        \node[punkt] (x) at (0.15,-0.5) {};
        \node[punkt] (y) at (1,-0.5) {} edge[<-] (x);
        \node[right] at (1.05,-0.5) {$\{\mathsf{U},\mathsf{E}\}$};
        \node[punkt] (x) at (0.15,-1) {};
        \node[punkt] (x) at (0.15,-1.5) {};
        \node[punkt] (x) at (0.15,-2) {};
      \end{scope}
      \end{scope}
    \end{tikzpicture}
    \caption{$G_0$ inspected through contexts.}
    \label{fig:graph-broken-down}
  \end{figure}
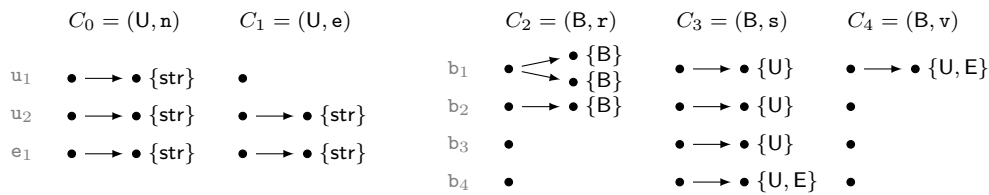
  Consider the type $\mathsf{U}$ with its 3 nodes in $G_0$: $\mathsf{u}_1$,
  $\mathsf{u}_2$, and $\mathsf{e}_1$, and fix the outgoing edge label to
  $\mathtt{n}$. In this context, which we denote $C_0=(\mathsf{U},\mathtt{n})$,
  each of the nodes has precisely one outgoing edge that leads to literal node
  of type $\str$. Naturally, when constructing a schema $S$ this should yield
  the corresponding type definition fragment $S(C_0)=\str^\ONE$. Analogously,
  for the context $C_1=(\mathsf{U},\mathtt{e})$ the corresponding fragment
  should be $S(C_1)=\str^\MAYBE$ since not every node has an outgoing
  $\mathtt{e}$-edge.

  Now, consider the type $\mathsf{B}$ and some of its contexts.  For the context
  $C_2=(\mathsf{B},\mathtt{r})$, we get quite naturally
  $S(C_2)=\mathsf{B}^\MANY$.  For $C_3=(\mathsf{B},\mathtt{s})$ we observe that
  in the graph $G_0$ whenever a node has type $\mathsf{E}$, it also has type
  $\mathsf{U}$, which indicates the type inclusion
  $\mathsf{E}\subseteq\mathsf{U}$. Consequently, the type fragment for $C_3$
  should be $S(C_3)=\mathsf{U}^\ONE$ rather than
  $\mathsf{U}^\MAYBE\mathsf{E}^\MAYBE$ for which there is insufficient evidence
  (such as a node of type $\mathsf{B}$ with two outgoing
  $\mathtt{s}$-edges). Finally, for $C_4=(\mathsf{B},\mathtt{v})$ two type
  definition fragments can be considered $\mathsf{U}^\MAYBE$ and
  $\mathsf{E}^\MAYBE$. Given the (scarce) evidence the reasonable choice seems
  $S(C_4)=\mathsf{E}^\MAYBE$ since the former option $\mathsf{U}^\MAYBE$ would
  be justified if there was a node of type $\mathsf{B}$ with an outgoing
  $\mathtt{v}$-edge leading to a node having the type $\mathsf{U}$ only. \qed
\end{example}

Formally, a \emph{context} is a pair $(t,a)\in\Gamma\times\Sigma$.  If
$\Gamma=\{t_1,\ldots,t_k\}$, a type definition \emph{fragment} is a string of
the form $t_1^{\mu_1}\ldots t_k^{\mu_k}$, where $\mu_i\in\M$ for
$1\leq i\leq k$. In the sequel, we abuse the notation and for a context
$C=(t,a)$ write $\arity_S^C(s)=\arity_S(t,a,s)$ if $(t,a,s)\in E_S$ and
$\arity_S^C(s)=\NONE$ otherwise. Then the type definition fragment corresponding
to a context $C$ in shape graph $S$ is
\[
S(C)=t_1^{\arity_S^C(t_1)}\ldots t_k^{\arity_S^C(t_k)}.
\] 
In the sequel, we denote by $\minarity_S$ and $\maxarity_S$ respectively the
minimum and the maximum value of $\arity_S$ respectively.

\paragraph{Inspecting graphs and graph languages} 
The previous example also shows that the information relevant to inferring a
given type definition fragment boils down to counting occurrences in the input
graph. Because the nodes of a typed graph are assigned sets of types, we first
identify all possible sets of types
\[
\Typesets(S) = \{\typing_G(n) \mid G\in L(S),\ n\in N_G\}.
\]
We point out that $\Typesets(s)$ can be constructed effectively by identifying
all sets of types that have nonempty intersection, which is know to be
decidable~\cite{StWi19}.

Now, for a context $C=(t,a)$, a graph $G$, and a typeset $T\in\Typesets(S)$ we
define the interval that contains the number of nodes having precisely types $T$
in the context $C$
\[
\occur_G^C(T) = \fit_\M(\{\, |\{m\in N_G \mid (n,a,m)\in E_G,\typing_G(m)=T\}| \, \mid n\in N_G,\ t\in\typing_G(n)\}).
\]
We extend the above construction to nonempty sets of typesets $\mathcal{T}\subseteq\Typesets(G)$
\[
  \occur_G^C(\mathcal{T}) = \fit_\M(\{\, |\{m\in N_G \mid (n,a,m)\in
  E_G,\typing_G(m)\in\mathcal{T}\}| \, \mid n\in N_G,\ t\in\typing_G(n)\}).
\]
Finally, for a type $t'$ we count its occurrences as 
\[
  \occur_G^C(t') = \fit_\M(\{\, |\{m\in N_G \mid (n,a,m)\in
  E_G,t'\in\typing_G(m)\}| \, \mid n\in N_G,\ t\in\typing_G(n)\}).
\]
In the sequel, we denote by $\minoccur_G^C(x)$ and $\maxoccur_G^C(x)$ the
minimal and the maximal value of $\occur_G^C(x)$. Also, we extend the above
notation to a language defined by a shape graph $S$ as
$\occur_{S}^C(x) = \occur_{\uplus L(S)}^C(x)$.  Naturally, our goal is to find
useful connections between $\occur$ that can be observed in examples and
$\arity$ that is in the shape graph. These connections can, however, be quite
involved.
\begin{example}[cont'd. Example~\ref{ex:contexts}]
  Take the schema $S_0$ in Figure~\ref{fig:example}. For the contexts that lead
  to a node with a single type, such as $C_0=(\mathsf{U},\mathsf{n})$,
  $C_1=(\mathsf{U},\mathsf{e})$, and $C_2=(\mathsf{B},\mathtt{r})$, the
  correspondence between $\occur$ and $\arity$ is straightforward e.g., 
  \begin{align*}
    &\occur_{S_0}^{C_0}(\mathsf{str})=\arity_{S_0}^{C_0}(\mathsf{str})=\ONE&
    &\occur_{S_0}^{C_1}(\mathsf{str})=\arity_{S_0}^{C_1}(\mathsf{str})=\MANY&
    &\occur_{S_0}^{C_2}(\mathsf{B})=\arity_{S_0}^{C_2}(\mathsf{B})=\MANY&
  \end{align*}
  For $C_3=(\mathsf{B},\mathtt{s})$ and $C_4=(\mathsf{B},\mathtt{v})$ which lead
  to nodes with possibly both types $\mathsf{U}$ and $\mathsf{E}$, their mutual
  relationship $\mathsf{E}\subseteq\mathsf{U}$ renders the connections between
  $\arity_S$ and $\occur_S$ far from obvious.
  \begin{align*}
    &
      \begin{aligned}
        &\arity_S^{C_3}(\mathsf{U})=\ONE \\
        &\arity_S^{C_3}(\mathsf{E})=\NONE 
      \end{aligned}
    &
    &
      \begin{aligned}
        & \occur_S^{C_3}(\{\mathsf{U}\})=\MAYBE\\
        & \occur_S^{C_3}(\{\mathsf{U},\mathsf{E}\})=\MAYBE\\
        & \occur_S^{C_3}(\{\mathsf{E}\})=\NONE
      \end{aligned}
    &
    &
      \begin{aligned}
        & \arity_S^{C_4}(\mathsf{U})=\NONE\\
        & \arity_S^{C_4}(\mathsf{E})=\MAYBE
      \end{aligned}
    &
    &
      \begin{aligned}
        & \occur_S^{C_4}(\{\mathsf{U}\})=\NONE\\
        & \occur_S^{C_4}(\{\mathsf{U},\mathsf{E}\})=\MAYBE\\
        & \occur_S^{C_4}(\{\mathsf{E}\})=\NONE
      \end{aligned}
    \tag*{\qed}
  \end{align*}
\end{example}

\paragraph{Connections between type occurrences and its arity in shape graph}
We now state a number of results that allow to establish connections between
occurrences of types in the input typed graph and the arities in the goal
schemas. Naturally, these connections are the basis of the work of our inference
algorithm.




First, we observe that type containment can be easily derived from a typed
graph.
\begin{proposition}
  For any two types $t_1$ and $t_2$ we have that $t_1\subseteq t_2$ if and only
  if for every $T\in\Typesets(S)$ we have $t_1 \in T \Rightarrow t_2 \in T$.
\end{proposition}

The appropriate connections between occurrences of a type and its arity in the
schema can be established in the presence of a sufficiently informative graph,
which we define next. 

\begin{definition}[Weakly characteristic graph]
  A typed graph $G$ is weakly characteristic of schema $S$ if the following
  conditions are satisfied:
  \begin{itemize}
    \itemsep0pt 
  \item for every $T\in\Typesets(S)$, there is a node in $G$ whose types are
    precisely $T$;
  \item for every context $C\in\Gamma\times\Sigma$ and every typeset
    $T\in\Typesets(S)$, there is a node that has exactly $\minarity^C_S(T)$ edges
    that goes to a node with types $T$, and another node that has exactly
    $\maxarity^C_S(T)$; if $\maxarity^C_S(T) = \infty$, we require at least
    $|T| + 1$ such edges.
  \end{itemize}
\end{definition}
We point out that its size may be exponential in the size of the goal schema and
the results on the sizes of counter-examples of containment of shape
graphs~\cite{StWi19} show that this bound is tight. 

We also point out that a weakly characteristic graph may contain insufficient
amount of information to infer the goal schema but it will allows us to
establish important links between $\occur$ and $\arity$. First we state the link
for the minimum values.
\begin{proposition}
  For a weakly characteristic graph $G$ for $S$, a context $C$, and a type $t$
  we have $\minoccur^C_G(t) = \sum\{\minarity^C_S(t') \mid t'\subseteq t \}$.
\end{proposition}
A less obvious link for maximum values is stated next. 
\begin{proposition}
  For a weakly characteristic graph $G$ for $S$ and a typeset $T\in\Typesets(S)$,
  $\maxoccur^C_G(T) \geq |T|+1$ if there is $t \in T$ with
  $\maxarity^C_S(t) = \infty$, and
  $\maxoccur^C_G(T) = \sum \{ \maxarity^C_G(t)\mid t\in T\}$ otherwise.
\end{proposition}
Finally, we point out important connection for equivalent types.
\begin{proposition}
  For any type $t$ take the set of equivalent types $T=\{t'\mid t=t'\}$. Then
  for any shape graph $S'$ that is equivalent to $S$, we have that
  $\sum_{s\in T}\minarity_{S}(s) = \sum_{s\in T} \minarity_{S'}(s)$
\end{proposition}


\section{Canonization of shape graphs}
\label{sec:canonization}
The connections between $\arity$ and $\occur$ are however more intricate than
even the above example suggests due to the fact that there might be many
equivalent shape graphs. 
\begin{example}
\label{ex:canon}
  In this example we consider 3 pairs of equivalent shape graphs presented in
  Figure~\ref{fig:canonization} and focus on a single context $C=(s,a)$.
  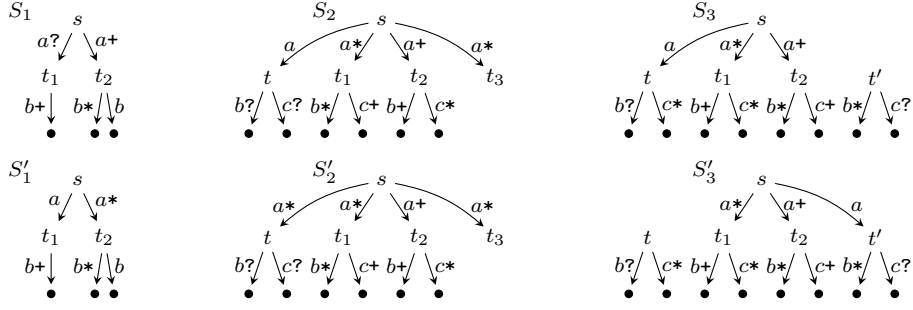
\begin{figure}[htb]
    \centering
\begin{tikzpicture}[
  >=stealth,
  punkt/.style={circle,minimum size=0.1cm,draw,fill,inner sep=0pt, outer sep=0.075cm}  
  ]
  \footnotesize
  \begin{scope}[xshift=0cm]
  \node at (-0.75,0.125) {$S_1$};
  \node (s) at (0,0) {$s$};
  \node (t1) at (-0.35,-0.75) {$t_1$} edge[<-] node[left,pos=0.75] {$a\MAYBE$} (s);
  \node (t2) at (0.35,-0.75) {$t_2$} edge[<-] node[right,pos=0.75] {$a\PLUS$} (s);
  \node[punkt] at (-0.35,-1.5) {} edge[<-] node[left=-1pt,pos=0.6] {$b\PLUS$} (t1);
  \node[punkt] at (0.225,-1.5) {} edge[<-] node[left=-1pt,pos=0.6] {$b\MANY$} (t2);
  \node[punkt] at (0.475,-1.5) {} edge[<-] node[right=-1pt,pos=0.6] {$b$} (t2);
  \end{scope}
  \begin{scope}[yshift=-2.125cm]
  \node at (-0.75,0.125) {$S_1'$};
  \node (s) at (0,0) {$s$};
  \node (t1) at (-0.35,-0.75) {$t_1$} edge[<-] node[left,pos=0.75] {$a$} (s);
  \node (t2) at (0.35,-0.75) {$t_2$} edge[<-] node[right,pos=0.75] {$a\MANY$} (s);
  \node[punkt] at (-0.35,-1.5) {} edge[<-] node[left=-1pt,pos=0.6] {$b\PLUS$} (t1);
  \node[punkt] at (0.225,-1.5) {} edge[<-] node[left=-1pt,pos=0.6] {$b\MANY$} (t2);
  \node[punkt] at (0.475,-1.5) {} edge[<-] node[right=-1pt,pos=0.6] {$b$} (t2);
  \end{scope}
  
  \begin{scope}[xshift=4cm]
  \begin{scope}[xshift=0cm, bend angle=15]
  \node at (-0.75,0.125) {$S_2$};
  \node (s) at (0,0) {$s$};
  \node (t) at (-1.5,-0.75) {$t$} edge[<-,bend left] node[left=1pt,pos=0.325] {$a$} (s);
  \node (t1) at (-0.5,-0.75) {$t_1$} edge[<-] node[left=-2pt,pos=0.6] {$a\MANY$} (s);
  \node (t2) at (0.5,-0.75) {$t_2$} edge[<-] node[right=-1pt,pos=0.6] {$a\PLUS$} (s);
  \node[inner sep=0.8pt] (t3) at (1.5,-0.75) {$t_3$} edge[<-,bend right] node[right=2pt,pos=0.3175] {$a\MANY$} (s);
  \node[punkt] at (-1.75,-1.5) {} edge[<-] node[left=-2pt,pos=0.6] {$b\MAYBE$} (t);
  \node[punkt] at (-1.25,-1.5) {} edge[<-] node[right=-1.25pt,pos=0.6] {$c\MAYBE$} (t);
  \node[punkt] at (-.75,-1.5) {} edge[<-] node[left=-2pt,pos=0.6] {$b\MANY$} (t1);
  \node[punkt] at (-.25,-1.5) {} edge[<-] node[right=-1pt,pos=0.6] {$c\PLUS$} (t1);
  \node[punkt] at (.25,-1.5) {} edge[<-] node[left=-2pt,pos=0.6] {$b\PLUS$} (t2);
  \node[punkt] at (.75,-1.5) {} edge[<-] node[right=-1pt,pos=0.6] {$c\MANY$} (t2);
  \end{scope}
  \begin{scope}[yshift=-2.125cm, bend angle=15]
  \node at (-0.75,0.125) {$S_2'$};
  \node (s) at (0,0) {$s$};
  \node (t) at (-1.5,-0.75) {$t$} edge[<-,bend left] node[left=1pt,pos=0.325] {$a\MANY$} (s);
  \node (t1) at (-0.5,-0.75) {$t_1$} edge[<-] node[left=-2pt,pos=0.6] {$a\MANY$} (s);
  \node (t2) at (0.5,-0.75) {$t_2$} edge[<-] node[right=-1pt,pos=0.6] {$a\PLUS$} (s);
  \node[inner sep=0.8pt] (t3) at (1.5,-0.75) {$t_3$} edge[<-,bend right] node[right=2pt,pos=0.3175] {$a\MANY$} (s);
  \node[punkt] at (-1.75,-1.5) {} edge[<-] node[left=-2pt,pos=0.6] {$b\MAYBE$} (t);
  \node[punkt] at (-1.25,-1.5) {} edge[<-] node[right=-1.25pt,pos=0.6] {$c\MAYBE$} (t);
  \node[punkt] at (-.75,-1.5) {} edge[<-] node[left=-2pt,pos=0.6] {$b\MANY$} (t1);
  \node[punkt] at (-.25,-1.5) {} edge[<-] node[right=-1pt,pos=0.6] {$c\PLUS$} (t1);
  \node[punkt] at (.25,-1.5) {} edge[<-] node[left=-2pt,pos=0.6] {$b\PLUS$} (t2);
  \node[punkt] at (.75,-1.5) {} edge[<-] node[right=-1pt,pos=0.6] {$c\MANY$} (t2);
  \end{scope}
  \end{scope}

  \begin{scope}[xshift=9cm]
  \begin{scope}[xshift=0cm, bend angle=15]
  \node at (-0.75,0.125) {$S_3$};
  \node (s) at (0,0) {$s$};
  \node (t) at (-1.5,-0.75) {$t$} edge[<-,bend left] node[left=1pt,pos=0.325] {$a$} (s);
  \node (t1) at (-0.5,-0.75) {$t_1$} edge[<-] node[left=-2pt,pos=0.6] {$a\MANY$} (s);
  \node (t2) at (0.5,-0.75) {$t_2$} edge[<-] node[right=-1pt,pos=0.6] {$a\PLUS$} (s);
  \node (t3) at (1.5,-0.75) {\phantom{$t$}} ;
  \node at (1.5,-0.75) {$t'$};
  \node[punkt] at (-1.75,-1.5) {} edge[<-] node[left=-2pt,pos=0.6] {$b\MAYBE$} (t);
  \node[punkt] at (-1.25,-1.5) {} edge[<-] node[right=-1.25pt,pos=0.6] {$c\MANY$} (t);
  \node[punkt] at (-.75,-1.5) {} edge[<-] node[left=-2pt,pos=0.6] {$b\PLUS$} (t1);
  \node[punkt] at (-.25,-1.5) {} edge[<-] node[right=-1pt,pos=0.6] {$c\MANY$} (t1);
  \node[punkt] at (.25,-1.5) {} edge[<-] node[left=-2pt,pos=0.6] {$b\MANY$} (t2);
  \node[punkt] at (.75,-1.5) {} edge[<-] node[right=-1pt,pos=0.6] {$c\PLUS$} (t2);
  \node[punkt] at (1.25,-1.5) {} edge[<-] node[left=-2pt,pos=0.6] {$b\MANY$} (t3);
  \node[punkt] at (1.75,-1.5) {} edge[<-] node[right=-1pt,pos=0.6] {$c\MAYBE$} (t3);
  \end{scope}
  \begin{scope}[yshift=-2.125cm, bend angle=15]
  \node at (-0.75,0.125) {$S_3'$};
  \node (s) at (0,0) {$s$};
  \node (t) at (-1.5,-0.75) {$t$} ;
  \node (t1) at (-0.5,-0.75) {$t_1$} edge[<-] node[left=-2pt,pos=0.6] {$a\MANY$} (s);
  \node (t2) at (0.5,-0.75) {$t_2$} edge[<-] node[right=-1pt,pos=0.6] {$a\PLUS$} (s);
  \node (t3) at (1.5,-0.75) {\phantom{$t$}} edge[<-,bend right] node[right=2pt,pos=0.3175] {$a$} (s);
  \node at (1.5,-0.75) {$t'$};
  \node[punkt] at (-1.75,-1.5) {} edge[<-] node[left=-2pt,pos=0.6] {$b\MAYBE$} (t);
  \node[punkt] at (-1.25,-1.5) {} edge[<-] node[right=-1.25pt,pos=0.6] {$c\MANY$} (t);
  \node[punkt] at (-.75,-1.5) {} edge[<-] node[left=-2pt,pos=0.6] {$b\PLUS$} (t1);
  \node[punkt] at (-.25,-1.5) {} edge[<-] node[right=-1pt,pos=0.6] {$c\MANY$} (t1);
  \node[punkt] at (.25,-1.5) {} edge[<-] node[left=-2pt,pos=0.6] {$b\MANY$} (t2);
  \node[punkt] at (.75,-1.5) {} edge[<-] node[right=-1pt,pos=0.6] {$c\PLUS$} (t2);
  \node[punkt] at (1.25,-1.5) {} edge[<-] node[left=-2pt,pos=0.6] {$b\MANY$} (t3);
  \node[punkt] at (1.75,-1.5) {} edge[<-] node[right=-1pt,pos=0.6] {$c\MAYBE$} (t3);
  \end{scope}
  \end{scope}

\end{tikzpicture}
\caption{Equivalent shape graphs: $S_1\equiv S_1'$, $S_2\equiv S_2'$, and
  $S_3\equiv S_3'$}
    \label{fig:canonization}
  \end{figure}
  For the schemas $S_1$ and $S_1'$ we observe that $t_1=t_2$, which gives the
  equivalence of the type definition fragments
  $S_1(C)=t_1^\MAYBE t_2^\PLUS = t_1^\ONE t_2^\MANY = S_1'(C)$. For the schemas
  $S_2$ and $S_2'$ we observe that $t\subseteq t_1\cup t_2\cup t_3$, which
  renders equivalent the fragments
  $S_2(C)=t^\ONE t_1^\MANY t_2^\PLUS t_3^\MANY= t^\MANY t_1^\MANY t_2^\PLUS
  t_3^\MANY=S_2'(C)$. Finally, we observe that
  $S_3(C)=t^\ONE t_1^\MANY t_2^\PLUS=t'^\ONE t_1^\MANY t_2^\PLUS = S_3'(C)$
  because $t\minus (t_1\cup t_2) = t' \minus (t_1\cup t_2)$.\qed
\end{example}

When a complete inference algorithm is presented with a characteristic graph of
a goal schema for which a number of equivalent formulations exists, a
well-behaved algorithm returns a formulation chosen according to clear
rules. These rules define a method of constructing a canonical shape graph that
we present next. Because this method needs to choose a single type among groups
of equivalent types, we facilitate this choice by fixing a total ordering $<$ of
the set of types $\Gamma$. Also, by $\minarity$ and $\maxarity$ we denote the
lower and upper bound of the interval of $\arity$, and we introduce similar
shortcuts for $\occur$.
\begin{definition}
  Take a shape graph $S\in\ShExZero$. We define the canonization operations of
  $S$ w.r.t.\ a context $C\in\Gamma\times\Sigma$
  \begin{description}
  \item[$(R1)$] if $t=t'$, $t < t'$, $\minarity(t)=0$, and
    $\minarity(t')=1$, then set $\minarity(t)=1$ and $\minarity(t')=0$;
  \item[$(R2)$] if $t\subseteq t_1\cup\ldots\cup t_k$ and $\maxarity(t_i)=\infty$
    for every $1\leq i \leq k$, then set $\maxarity(t)=\infty$;
  \item[$(R3)$] if
    $t\setminus(t_1\cup\ldots\cup t_k) = t'\setminus(t_1\cup\ldots\cup t_k)$,
    $t < t'$, $\arity(t')=\NONE$, $\arity(t)=\MAYBE$, and
    $\maxarity(t_i)=\infty$ for every $1\leq i \leq k$, then set
    $\arity(t)=\NONE$ and $\arity(t')=\MAYBE$.
  \end{description}
  By $\Can^<(S)$ we denote the shape graph obtained by applying exhaustively the
  rule (R1) in every context, then exhaustively the rule (R2) in every context,
  and finally, exhaustively the rule (R3) in every context. We say that $S$ is
  \emph{canonical} w.r.t.\ $<$ iff $\Can^<(S)=S$ 
\end{definition}

\begin{example}[cont'd. Example~\ref{ex:canon}]
  $S_1'$ is obtained from $S_1$ by applying the rule (R1) since $t_1$ and $t_2$
  are equivalent. $S_2'$ is obtained from $S_2$ by applying the rule (R2)
  because $t$ is covered by $t_1$, $t_2$, and $t_3$ i.e.,
  $t\subseteq t_1\cup t_2\cup t_3$.  Finally, $S_3'$ is obtained from $S_3$ by
  applying the rule (R3) because
  $t\minus (t_1\cup t_2) = t' \minus (t_1\cup t_2)$. \qed

\end{example}
We next state and prove the main result of this section.  
\begin{theorem}
  For any two $S\equiv S'$ we have $\Can^<(S)=\Can^<(S')$.
\end{theorem}
Below, we outline the proof of the above theorem and we assume a fixed order $<$
on types, fix two schemas $S$ and $S'$, assume that they are equivalent
$S\equiv S'$, and let $S_1=\Can^<(S)$ and $S_2=\Can^<(S')$. It is relatively
straightforward to show that each of the canonization operations preserves the
semantics.
\begin{claim}
$S_1\equiv S\equiv S'\equiv S_2$.
\end{claim}
We next show that the exhaustive application of the rule (R1) ensures equality of
the minimums of each arity.
\begin{claim}
  For any context $C$ and any type $t$ we have
  $\minarity_{S_1}^C(t)=\minarity_{S_2}^C(t)$.
\end{claim}
The exhaustive application of the rule (R2) ensures that each infinite maximum
arity is the same.
\begin{claim}
  For any context $C$ and any type $t$ we have $\maxarity_{S_1}^C(t)=\infty$ iff
  $\maxarity_{S_2}^C(t)=\infty$
\end{claim}
Finally, the exhaustive application of the rule (R3) guarantees that the
multiplicities $\NONE$, $\MAYBE$, and $\ONE$ are the same. 
\begin{claim}
  For any context $C$ and any type $t$ if $\maxarity_{S_1}^C(t)<\infty$ or
  $\maxarity_{S_2}^C(t)=\infty$, then 
  $\maxarity_{S_1}^C(t) = \maxarity_{S_2}^C(t)=\infty$.
\end{claim}


\section{Inference of Shape Graphs}
\label{sec:infererence}
We now present the inference algorithm for typed graphs. Because of space
restrictions, we only present its outline on the graph in
Figure~\ref{fig:graphs} and the contexts in Example~\ref{ex:contexts}. The
detailed algorithm can be found in appendix. 

\noindent
For a given graph $G$ the algorithm $\typedlearner$ performs the following
steps.
\begin{enumerate}
\item It begins by gathering the typesets present in the input graph
  $\mathcal{T}=\{types_G(n) \mid n\in N_G\}$. \emph{In our example}
  \[
    \mathcal{T} = \{\{\mathsf{B}\},\{\mathsf{U}\},
    \{\mathsf{U},\mathsf{E}\},\{\str\}\}
  \]
\item It uses the existing evidence to establish the inclusion relationship between
  the types $t\subseteq s$ is assumed to hold iff for every $T\in\mathcal{T}$ we
  have that $t\in T$ implies $s\in T$. \emph{In our example}
  \[
    \mathsf{\str}\subseteq\mathsf{\str},\;
    \mathsf{E}\subseteq\mathsf{E},\;
    \mathsf{E}\subseteq\mathsf{U},\;
    \mathsf{U}\subseteq\mathsf{U},\;
    \mathsf{B}\subseteq\mathsf{B}.\;
  \]
\item it fixes an order $\lhd$ of enumerating the types $\Gamma$ that
  is compatible with $<$ and $\subseteq$: $t\lhd s$ whenever
  $t\subseteq s$ or if $t$ and $s$ are incomparable by $\subseteq$ and $t<s$. \emph{In
  our example we set}
  \[
    \mathsf{\str}\lhd
    \mathsf{E}\lhd
    \mathsf{U}\lhd
    \mathsf{B}
  \]
\item For every context $C$ it enumerates the types used in this context in the
  order $\lhd$ and infers the minimal arities:
  $\minarity(t)=\minoccur(t)-\sum_{s\subseteq t}\minarity(s)$. \emph{In our example}
  \begin{align*}
    &\minarity^{C_2}_{G_0}(\mathsf{B}) = 0, & 
    &\minarity^{C_3}_{G_0}(\mathsf{E}) = 0, & 
    &\minarity^{C_3}_{G_0}(\mathsf{U}) = 1, \\
    &&
    &\minarity^{C_4}_{G_0}(\mathsf{E}) = 0, &
    &\minarity^{C_4}_{G_0}(\mathsf{U}) = 0. 
  \end{align*}

\item For every context $C$ it enumerates the types used in this context in the
  reversed order $\lhd^{-1}$ and infers the maximal arity according to
  one of the following cases:
  \begin{enumerate}
  \item If $\maxoccur(T)\geq |T|+1$ for all typesets $T\in\mathcal{T}$
    containing $t$, then it sets $\maxarity(t)=\infty$; \emph{In our example, this
    applies to $\maxarity^{C_2}_{G_0}(\mathsf{B}) = \infty$}
  \item Otherwise, if $\minarity(t)=1$, then it sets $\maxarity(t)=1$; \emph{In our
    example, this applies to $\minarity^{C_3}_{G_0}(\mathsf{U}) = 1$}
  \item Otherwise, it looks for a typeset $T$ that \emph{characterizes} $t$ for
    $C$, in the sense that $T$ contains only $t$ and types $s$ with $t \subseteq s$, 
    and sets $\maxarity(t)=\maxoccur(T)-\sum_{s \in
      T\setminus \{t\} }\maxarity(s)$. \emph{In our example,}
    \begin{align*}
      &\maxarity^{C_3}_{G_0}(\mathsf{E}) = 0&&\text{because $\{\mathsf{U},\mathsf{E}\}$ characterizes $\mathsf{E}$ for $C_3$,}\\
      &\maxarity^{C_4}_{G_0}(\mathsf{U}) = 0& & \text{because $\{\mathsf{U}\}$ characterizes $\mathsf{U}$ for $C_4$,}\\
      &\maxarity^{C_4}_{G_0}(\mathsf{E}) = 1& & \text{because $\{\mathsf{U},\mathsf{E}\}$ characterizes $\mathsf{E}$ for $C_4$.}
    \end{align*}
  \item If there does not exist a typeset $T$ that characterizes $t$ (typically
    when $t$ is a union of other types), $t$ is said to be \emph{obfuscated} for
    $C$.  In which case, we look for typesets $T_i$ that contains $t$ and that
    have $\maxarity^C_S(t) \neq \inf$ for all $t \in T_i$.
    
    If there is only one such $T_i$, then we set $\maxarity^C_G(t) = 1$ if
    \[
      \maxoccur^C_G(T) - \sum\{\maxarity^C_G(t') \mid t' \in T,\ t' > t\}  - 
      \sum \{\minarity^C_G(t') > 0 \mid t' \in T, t' < t\},
    \] 
    and we set $\maxarity^C_G(t) = 0$ otherwise.
 	
    If there are more than one such $T_i$, we pick any two typesets $T_1$ and
    $T_2$. We compute $N = \Sigma_{ t'\in T \setminus \{t\}} \maxarity^C_S(t')$.
    We set $\maxarity^C_S(t) = 0$ if
    $\maxoccur^C_S(\{T_1, T_2 \}) - N = \Sigma_i (\maxoccur^C_S(T_i) - N)$,
    otherwise, we set $\maxarity^C_S(t) = 1$
  \end{enumerate}
\item Finally, the result may not be consistent with the input graph, in
  particular if the graph is not characteristic. The algorithm therefore has to
  check consistency for all type definitions, and relax locally all type
  definitions which are not consistent.
\end{enumerate}

The notion of characteristic graph $G$ for shape graph $S$ is central. As
explained before, we require the graph $G$ to be weakly characteristic for
$S$. This will allow to give enough information to infer the canonical
representative of $S$ unless $S$ has obfuscated types.


If there are obfuscated types, we need to add extra information to the
characteristic graph. For every context $C$, and type $t$ that is obfuscated for
$C$, we consider all typesets $T_i$ that contain $t$ and such that
$\maxarity^C_S(t) \neq \infty$. We denote the set of such typesets by
$\cover(t)$. Now, a graph $G$ is \emph{characteristic} for $S$ if it is weakly
characteristic for $S$ and if for every type $t$ obfuscated for a context
$C=(t_0,a)$ with $|cover(t)| \geq 2$, for every pair
$\{ T_1, T_2\} \subset cover(t)$, there is a node $n$ of type $t_0$ that such
that the number of $a$-labelled edges that go to a node that has $T_1$ or $T_2$
as a typing is maximal, i.e. is equal to
$\Sigma_{ t' \mid t' \in T_1 \cup T_2 } \maxarity^C_S(t')$.

\begin{theorem}
  $\ShExZero$ is learnable in polynomial time from typed graphs.
\end{theorem}
We claim that if the input graph is characteristic for a shape graph $S$, the
algorithm outputs $\Can(S)$. We outline the proof of the completeness below.
\begin{enumerate}
\itemsep 0pt
\item The algorithm gathers the correct typeset as each typeset that may exist
  is present in the graph.
\item The algorithm computes the correct inclusion relation between types,
  because $s \subseteq t$ implies that $(t \in T \Rightarrow s \in T)$ for any
  typeset $T$, and if $s \not \subseteq t$, then there exists a typeset $T$ that
  contains $t$ but not $s$, and this typeset is present in a characteristic
  graph.

\item Minimal arities are computed correctly. This is proved recursively: for
  minimal types (i.e. types that contains no other types), we directly have
  $\minoccur^G_S(t) = \minarity^G_S(t)$.  For larger types,
  $\minoccur^C_G(t) = \minarity^C_S(t) + \Sigma_{s \subseteq t}
  \minarity^C_S(s)$, and hence
  $\minarity^C_S(t) = \minoccur^C_G(t) - \Sigma_{s \subseteq t}
  \minarity^C_S(s)$.

\item For maximal arities, there are four cases.
\begin{enumerate}
\item If $\maxarity^G_S(t) = \infty$, then evidence of it can be found in a
  characteristic graph. Indeed, for a typeset $T$, if
  $\maxoccur^C_G(T) \geq |T|+1$, that means that there is a type $t \in T$ for
  which $\maxarity^C_S(t) = \infty$. For a type $t$, if we consider a typeset
  $T$ that contains only smaller types $s \subset t$, then either $t$ of some
  $s \in T \setminus \{t\}$ has an infinite maximal arity. But in a canonical
  shape graph $S$, if this happens for all $T$ that contain $t$, then also
  $\maxarity^C_S(t) = \infty$.

\item Other case is when $\maxarity^C_S(t) \neq \inf$. Then, if
  $\minarity^C_S(t) = 1$, then it implies that $\maxarity^G_S(t) = 1$.

\item It remains only cases where $\arity^C_S(t) \in\{\NONE,\MAYBE\}$. If there
  is a typeset $T$ that contains only $t$ and its super-types, we can do as for
  minimal arities and use the fact that
  $\maxoccur^C_G(t) = \maxarity^C_S(t) + \sum\{\maxarity^C_S(s)\mid s \supseteq
  t\}$ (note that all $s$ have finite arity).

\item If such a typeset $T$ does not exist, the $t$ is obfuscated for $C$
  Consider all typesets $T_i$ that contain $t$ and that have
  $\maxarity^C_S(\{T_i\}) \neq \infty$. Note that there has to be at least one
  $T_i$, otherwise rule (R2) could be applied and then
  $\maxarity^C_S(t) = \inf$.

  If there is only one such $T_i$ (say $T_1$), rule (R2) can be applied, and we
  put the highest possible maximal arity on $t$. We compute
  $\maxoccur^C_G(T_1)$, and remove $\maxarity^C_S(t')$ for $t' \in T$ and
  $t' > t$ (they are already computed), and remove $\minarity^C_S(t')$ for
  $t' \in T$ and $t' > t$ (as $\maxarity^C_S(t')$ can't be lower than
  $\minarity^C_S(t')$ and these are also already computed).

  We next consider the case where there are several such $T_i$'s. We pick two,
  say $T_1$ and $T_2$. For instance, consider four types $t, t_0, t_1, t_2$ with
  $t \subseteq t_0$ and $t = t_1 \cup t_2$. For a context $C$, imagine the
  target arity is $\arity^C_S(t_0) = 1$, $\arity^C_S(t) = ?$,
  $\arity^C_S(t_1) = 1$ and $\arity^C_S(t_2) = 0$.  We then consider
  $T_1 = \{ t_0, t, t_1 \}$ and $T_2 = \{ t_0, t, t_2 \}$.

  The general case is similar: the type $t$ may be included in larger types
  ($t_0$ here), but with $\maxarity(t) \neq \infty$ (otherwise
  $\maxarity(t) = \inf$ as well).  Note that $\maxarity^C_S(t_0)$ is already
  computed since $t < t_0$. Similarly, there are at least two types $t_1$ and
  $t_2$ that have nonempty intersection with $t$ and for which $\maxarity$ is
  not $\infty$ (other cases are already treated before).

  We compute $N = \sum\{\maxarity^C_S(t')\mid t' \in T \setminus \{t\} \}$.  In
  our example, we have $N = \maxarity^C_S(t_0) = 1$. It is of course possible
  that $N=0$ if there is no super-type for $t$.  We have then
  $\maxoccur^C_G(\{T_1, T_2\}) = 3$, $\maxoccur^C_G(\{T_1\}) = 3$, and
  $\maxoccur^C_G(\{T_2\}) = 2$. If we add up the elements
  $\maxoccur^C_G(\{T_1, T_2\}) - N = 2$ whereas
  $(\maxoccur^C_G(\{T_1\}) -N) + (\maxoccur^C_G(\{T_2\}) -N) = 2 + 1 = 3$. This
  means that $t$ should be counted both in $T_1$ and $T_2$, and so
  $\maxarity^C_S(t) = 1$.
\end{enumerate}
\end{enumerate}


\section{Related work}
\label{sec:related}


There exists a host of research on schemas for semistructured (graph) data
\cite{BunemanDFS97,fernandez1998optimizing} and their inference
\cite{GoldmanW97, NestorovUWC97, NestorovAM98}. Dataguides \cite{GoldmanW97} are
data structures that represent all paths in semistructured graph but it is
assumed that graph have entry points, essentially root nodes, which is not an
assumption we make for RDF graphs. \cite{NestorovAM98} the typing is by a
datalog program, the typings allow objects to have multiple types at the same
time which is essential when the data is fairly irregular. Approximate merging
of types using clustering algorithms is applied until the typing is of
acceptable size.

Several approaches have been proposed for inference of schemas for
tree-structured data \cite{garofalakis2003xtract,min2003efficient,
  chidlovskii2001schema,bex2006inference,bex2007inferring,BaaziziLCGS17,BaaziziCGS19}. XTract
\cite{garofalakis2003xtract} infers DTD schemas by generating candidate regular
expressions for each element name and then selecting the best one. Another
method is to generate various versions of finite automata (probabilistic,
Glushkov, etc.) and then rewrite them into regular expressions
\cite{bex2006inference, chidlovskii2001schema}.  Inference of more expressive
schema formalisms such as unranked tree automata or XML Schema is considered in
\cite{bex2007inferring, carme2004learning,
  raeymaekers2005learning,fernau2002learning}. These approaches also include inference of schema for JSON data \cite{BaaziziLCGS17,BaaziziCGS19} that by inferring types of each node by merging the types of their children in a manner analogous to determinization. 


Another relevant areas of research is graph summarization, where the goal is to
compute compact but accurate representation of the input graph.  One approach is
to compute quotient graph w.r.t. an equivalence relation by collapsing nodes in
the same equivalence class into one.  Many applications of this technique are
based on (bi)simulation \cite{fan2012query,zhang2014assg,
  khatchadourian2010explod,tran2013managing}, often taking into account
particular queries that are to be answered. In
\cite{Goasdoue19,CebiricGM17,CebiricGGM18}) the author also propose a
quotient-based method but introduce two equivalence relations that are not based
on bisimulation. The first relation, \emph{strong equivalence}, requires
similarity of the structure of incoming and outgoing edges, and \emph{weak
  equivalence}, which requires similarity between incoming or outgoing edges. We
point out that often the approaches make the assumption of one type per node,
which simplifies the type definition process. \cite{GoasdoueGM20} follows this
line of research presenting four other summaries that are representable in the
style of E-R diagrams together with efficient algorithms for their computation.
In \cite{MadukoASS08} graph summaries are computed based on estimating the
frequency with which subgraphs match given query patterns.  The paper
\cite{ArenasDFKS14} presents an approach where users first define a
\emph{structuredness} function $\sigma$ that measures how well a given RDF graph
fits to the schema and then discover a partitioning of the entities of an RDF
graph into subsets which have high structuredness. They consider an optimization
variant of the inference problem: finding the lowest number of types for a given
threshold on $\sigma$ or finding a fixed number of types that maximizes
$\sigma$. This approach as most others can be described as \emph{pragmatic}, the
goal is to find a schema that is as small as possible and describes the data
with a good precision.  Our motivations, in contrast, are more fundamental and
aim at understanding the inherent limitations of inference.

\vspace*{-10pt}
\section{Conclusions and future work}
\label{sec:conclusions}
In the present paper we have studied the problem of inference of shape
expression schemas for typed RDF graphs. We have presented a sound and complete
inference algorithm for a practical subclass \ShExZero of shape expressions
schemas. Our investigation lead us to study the canonization problem for
\ShExZero and we present an effective canonization procedure for \ShExZero.


\bibliographystyle{abbrv}
\bibliography{staworko,paper}

\end{document}